\documentclass[]{llncs}
\usepackage[margin=2.5cm]{geometry}
\usepackage{graphicx}
\usepackage{fancyvrb,courier}
\usepackage{url}
\usepackage{amsmath}
\usepackage{amsfonts}
\usepackage{amssymb}

\begin{document}
\mainmatter
\title{Automated Verification of Quantum Protocols\\ by\\ Equivalence Checking}
\author{Ebrahim Ardeshir-Larijani\inst{1} }
\institute{Department of Computer Science\\University of Warwick\\\mailsb}

\author{Ebrahim Ardeshir-Larijani\inst{1,2}\thanks{Supported by the
    Centre for Discrete Mathematics and its Applications (DIMAP),
    University of Warwick, EPSRC award EP/D063191/1.} \and Simon J.\ Gay\inst{2}  \and Rajagopal Nagarajan\inst{3}\thanks{
Partially supported by ``Process Algebra Approach to Distributed Quantum Computation and Secure Quantum Communication", Australian Research Council Discovery Project DP110103473.}}
\institute{Department of Computer Science, University of
  Warwick\\\mailsb \and School of Computing Science, University of
  Glasgow\\\mailsa \and Department of Computer Science,\\
  School of Science and Technology, Middlesex University \\\mailsc}
\urldef{\mailsa}\path|Simon.Gay@glasgow.ac.uk|
\urldef{\mailsb}\path|E.Ardeshir-Larijani@warwick.ac.uk|
\urldef{\mailsc}\path|R.Nagarajan@mdx.ac.uk|
\maketitle
% importance of formal methods in current/classical technologies, giving understanding ranging from 
%software/hardware systems to system biology.
%Model checking along examples of industerial use and Equivalence Checking

%Current work / language / semantics / experiemental results
%-------------------------------------------------------------------------------------------------
\section{Prologue}
Technologies based on \emph{Quantum Information Processing (QIP)} are emerging rapidly in our life, from cryptography and communication
to fast computation.
The grand challenge in QIP is not only harnessing and controlling natural processes but is also about 
how we deploy QIP to solve our problems. This is where \emph{conceptual errors} in designing  
QIP protocols may arise.
This is because the correctness of QIP protocols relies on theoretical proofs on the paper, while sophisticated 
QIP implementations demand systematic and automated verification.
\emph{Classical Information Processing (CIP)} also shares the same problem, and that is why
a range of techniques and tools, called \emph{formal verification} have been developed to verify large and interactive systems such as systems on chip, communication and cryptography protocols,
against design errors. 
Formal verification deals with \emph{mathematical models} of systems and abstracts their requirements by specifying
them in formal languages, and systematically checks that whether those requirements are satisfied in the 
system's model.
The goal of our novel, ongoing research is to develop formal verification techniques and tools for analysis of QIP protocols.
\section{Formal Verification}
Formal verification has evolved tremendously in the past few decades from a conceptual framework for analysing software and hardware systems 
to successful industrial practices such as verifying safety-critical and cryptographic systems. Different methods and tools have been developed for the verification of systems,
resulting in more reliable and safer computer systems. One example of such method is \emph{Model Checking} \cite{pmc}.
The aim of model checking is to explore the behaviour of a system \emph{exhaustively}, in an attempt to find errors.
The outcome of such an exhaustive search is either a verified model of the system or a counterexample that tells us where things
went wrong. Thus we have following steps in model checking of systems: (1) \emph{representing} a model of a given system in a formal language
(e.g.\, programming languages like C, Java, etc.); (2) constructing a \emph{model} i.e.\, 
preprocessing and interpreting the input representation of the model, making it usable for automated reasoning.
The challenge in model checking is that often the generated model has a huge size, making automated reasoning difficult; 
 (3) specifying requirements of the system, normally as a logical formula; (4) checking the formula
on the generated \emph{model}; (5) representing the outcome, either as satisfaction of requirements (\emph{specification}) or providing traces of the system where requirements are not satisfied (counter examples).      
\section{Equivalence Checking for Quantum Systems}
\begin{figure}[hbtp]
\vspace{-7mm}
\begin{center}
\includegraphics[height=6 cm,width=12cm]{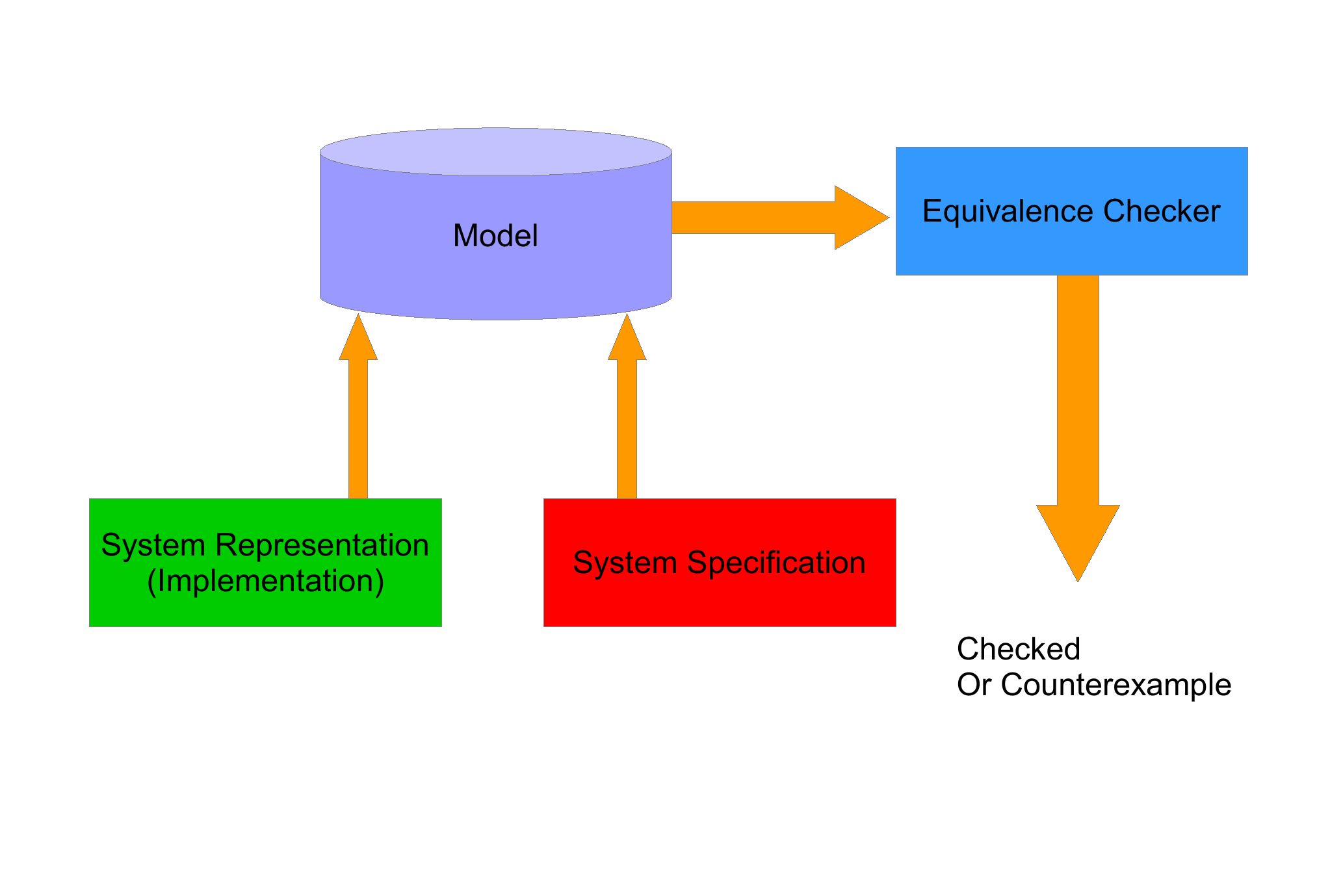}
\end{center}
\vspace{-10mm}
\caption{Equivalence Checking}
\label{fig:ec}
\end{figure}

In \cite{qmc-original} and \cite{chapqmc} Gay, Nagarajan and Papanikolaou have developed a \emph{Quantum Model Checker (QMC)} for checking models of quantum protocols such as Teleportation with \emph{stabilizer} states as input. 
QMC verifies protocols within stabilizer formalism on given stabilizer states as input.
Nevertheless, the results of QMC can be interpreted as   
an \emph{evidence} for the correctness and not the \emph{proof} of correctness
of quantum protocols.
In this work, however, we verify such protocols for 
arbitrary input states, using \emph{linearity} of superoperators, in order to prove the correctness. We have implemented a  variant of model checking,
called \emph{equivalence checking}. The aim here is to show two representations of protocols, one corresponding to the specification and another to the implementation, are equivalent, proving the implemented protocol satisfies the specification.
The schematic process of equivalence checking is illustrated in Figure~\ref{fig:ec}.
Our tool 
\emph{Quantum Equivalence Checker (QEC)} follows the steps described in Figure~\ref{fig:ec}, namely:
to represent models, we have designed a language in which processes are defined separately, capable of communicating classical and
quantum data. This is, of course, much more expressive than quantum circuit languages.
Note that similar languages for analysing concurrent QIP systems have been developed, but they lack tool support
(see~\cite{cqp}, \cite{qccs}).
Our language is inspired by the influential work of
Milner \cite{mccs}, in \emph{concurrency} theory.
The core idea of Milner's formalism is that the behaviour of a concurrent system is understood based on what is 
\emph{observable} from outside and how different processes interact with each other. Another reason for choosing this
formalism is \emph{synchronised} communication, given the lack of durable quantum memory.
In this language, processes are formed using
prefixes, $``."$, and parallel composition $``|"$. The simplest process is $nil$, with no action.
Also quantum operations are treated as prefixes.
For example, $Y(a).nil$ or $X(a).nil\ |\ Z(b).nil$,
where $X$,$Y$ and $Z$ are Pauli operators, form instances of terms in this language. 
Sending and receiving a bit/qubit
$x$ over a channel 
$c$ is done by $c!x$ and $c?x$, respectively. As an example of using this language, Teleportation and
its specification is illustrated in Figure~\ref{fig:lang}. In contrast to quantum circuit diagrams, explicit communications between separate parties 
and parallel compositions can be expressed. This makes clear how parties interact with each other.
\begin{figure}
\vspace{-5mm}
\begin{Verbatim}[frame=single,fontsize=\scriptsize]
Implementation =
//Preparing EPR pair and sending to Alice and Bob:
newqubit y . newqubit z . H(y) . CNOT(y,z) . c!y . d!z . nil
|

//Alice's process:
input x . c?y . CNOT(x,y) . H(x) . m := measure x . n := measure y. b!m . b!n . nil

|

//Bob's process :
d?w . b?m . b?n . if n then X(w) . if m then Z(w) . output w . nil
\end{Verbatim}
\begin{Verbatim}[frame=single,fontsize=\scriptsize]

Specification = input x.output x.nil
\end{Verbatim}
\vspace{-5mm}
\caption{Concurrent Teleportation and its Specification}
\label{fig:lang}
\end{figure}
A model is built by \emph{scheduling} concurrent processes and \emph{simulating} in the \emph{stabilizer formalism}.
For checking equivalence, we have defined a formal \emph{semantics} of our model in terms of superoperators. This is inspired by the 
Quantum Programing Language in \cite{qpl}. 
Now we want to check that implementation of a protocol behaves equivalently to its specification, 
e.g.\, $\textit{Implementation}\simeq \textit{Specification}$, that is to prove the two 
superoperators, corresponding to implementation and specification are equal.
We check the equivalence of superoperators by using linearity, i.e.\ computing the effect of superoperators on 
the basis density operators and test the equality of final states.    
For testing equality of stabilizer states, we have designed an algorithm that checks
the linear independence of stabilizer generators, partly using efficient algorithms in \cite{aden}.
We have done a range of experiments on useful quantum protocols, ranging from Teleportation to Quantum Secret Sharing. We have used both sequential
(models without parallel composition and communications) and concurrent
models. In concurrent models, a sequence of actions are called interleaving, whereas in sequential models, they are known as branching. The experimental results are shown in the Figure~\ref{tb:experiments}. 
\begin{figure}
\begin{center}
\begin{tabular}{|c|c|c||c|c|c|}
\hline
\textbf{Protocol}  &\tiny{No. Interleaving} &\tiny{CM}&\tiny{No. Branch}&\tiny{SM} & \tiny{SEC} \\ \hline
Teleportation &  400 & 343 &16 &39& 43 \\ \hline
Dense Coding &   100 & 120& 4&22 &30\\ \hline
Bit flip code &  16 & 62 &16 &60 &61\\ \hline
Phase flip code &   16 & 63 &16 &61& 62 \\ \hline
Five qubit code &   64 & 500 & 64&451 &*\\ \hline
X-Teleportation &   32 & 63  & 8&18 &25\\ \hline
Z-Teleportation &   72 & 78  & 8& 19&27\\ \hline
Remote CNOT &   78400 & 12074 &64 &112 & 140\\ \hline
Remote CNOT(A) &   23040 & 4882 & 64&123 &156 \\ \hline
Quantum Secret Sharing &   88480 & 13900& 32&46 &60 \\ \hline
\end{tabular}
\end{center}
\vspace{-5mm}
\caption{\scriptsize{Experimental results of equivalence checking of
    quantum protocols.  The columns headed by CM and SM show the
    results of verification of concurrent and sequential models of
    protocols in our current tool. Column SEC shows verification times
    for sequential models in our previous tool \cite{tacas}. The
    number of branches for SM and SEC models are the same.
    The model for (*) is not available due to the tediousness of implementation.     
     Times are
    in milliseconds.}}
\label{tb:experiments}
\end{figure}        
\section{Conclusion}
We have presented an equivalence checking technique for automated verification of QIP protocols.
We have used a more expressive language than circuit diagrams to specify concurrent QIP protocols and furthermore,
we have implemented the tool QEC to carry out experimental results. Extending our technique beyond the stabilizer formalism and also
developing more general notions of equivalence, such as \emph{bisimulation} (similar to the approach of~\cite{symbis}), will be our future directions in this research.  
\bibliographystyle{simonplain}
\bibliography{paper}
\end{document}